\documentclass[11pt]{article}
\usepackage{axodraw}
\usepackage{epsfig}
\usepackage{amsfonts}
\usepackage{amsmath}
\usepackage{bbm}
\input pix.sty

\renewcommand{\eq}{eq.~}
\renewcommand{\eqs}{eqs.~}

\newcommand{\mD}{m_\rmi{D}}

\newcommand{\Nc}{N_{\rm c}}

\newcommand{\rmO}{{\mathcal{O}}}

\def\lsi{\raise0.3ex\hbox{$<$\kern-0.75em\raise-1.1ex\hbox{$\sim$}}}
\def\gsi{\raise0.3ex\hbox{$>$\kern-0.75em\raise-1.1ex\hbox{$\sim$}}}

\newcommand{\nB}{n_\rmii{B}}

 \renewcommand{\nB}[1]{n_\rmii{B{#1}}}
\newcommand{\rmii}[1]{{\mbox{\tiny\rm{#1}}}}
\newcommand{\re}{\mathop{\mbox{Re}}}
\newcommand{\im}{\mathop{\mbox{Im}}}

\newcommand{\Tint}[1]{{\hbox{$\sum$}\!\!\!\!\!\!\!\int\,}_{\!\!\!\!\raise-0.9ex\hbox{$\scriptstyle{#1}$}}}
\newcommand{\Tinti}[1]{{{\Sigma}\!\!\!\!\raise0.3ex\hbox{$\int$}_\rmii{${#1}$}}}

\newcommand{\bi}{\begin{itemize}}
\newcommand{\ei}{\end{itemize}}

\newcommand{\hide}[1]{ }

\def\TAsc(#1,#2)(#3,#4,#5)%
{\SetWidth{2.0}\CArc(#1,#2)(#3,#4,#5)\SetWidth{1.0}}
\def\Lwidth{3}

\def\TAgl(#1,#2)(#3,#4,#5){\SetWidth{2.0}\PhotonArc(#1,#2)(#3,#4,#5){\Lwidth}%
{6.283 #3 mul 360 div #4 #5 sub #4 #5 sub mul sqrt mul Tdensity mul}%
\SetWidth{1.0}}
\def\TLgl(#1,#2)(#3,#4){\SetWidth{2.0}\Photon(#1,#2)(#3,#4){\Lwidth}
{#1 #3 sub #1 #3 sub mul #2 #4 sub #2 #4 sub mul add sqrt Tdensity mul}%
\SetWidth{1.0}}
\newcommand{\piC}[1]{\;\parbox[c]{40pt}{\begin{picture}(120,60)(0,-20)
\SetWidth{1.0}\SetScale{0.35} #1 \end{picture}}\;}
\def\ConnectedA(#1,#2,#3){\piC{#1(60,-15)(75,34,146) #2(60,75)(75,214,326)%
 #3(60,60)(20,190,350)%
 \GBoxc(0,30)(10,10){1} \GBoxc(120,30)(10,10){1}%
  }}
\def\ConnectedB(#1,#2,#3){\piC{#1(60,-15)(75,34,146) #2(60,75)(75,214,326)%
 #3(60,60)(60,0)%
 \GBoxc(0,30)(10,10){1} \GBoxc(120,30)(10,10){1}%
  }}
\def\ConnectedC(#1,#2){\piC{#1(60,-15)(75,34,146) #2(60,75)(75,214,326)%
 \GBoxc(0,30)(10,10){1} \GBoxc(120,30)(10,10){1}%
  }}
\def\ConnectedD(#1,#2){\piC{#1(60,-15)(75,34,146) #2(60,75)(75,214,326)%
 \GBoxc(0,30)(10,10){1} \GBoxc(120,30)(10,10){1}%
 \SetWidth{2.0} 
 \Line(55,55)(65,65)%
 \Line(55,65)(65,55)
  }}

\begin{document}

\begin{titlepage}
\begin{flushright}
BI-TP 2009/09\\
\vspace*{1cm}
\end{flushright}
\begin{centering}
\vfill

{\Large{\bf
Quarkonium dissociation in the presence of \\[2mm] 
a small momentum space anisotropy
}} 

\vspace{0.8cm}

Y.~Burnier, 
M.~Laine, 
M.~Veps\"al\"ainen 

\vspace{0.8cm}

{\em
Faculty of Physics, University of Bielefeld, 
D-33501 Bielefeld, Germany\\}

\vspace*{0.8cm}

\mbox{\bf Abstract}
 
\end{centering}

\vspace*{0.3cm}
 
\noindent
We consider the dissociation of heavy quarkonium 
in a medium close to thermal equilibrium but with 
a small momentum space anisotropy. Dissociation is 
defined to take place when the width of the ground 
state equals its binding energy. We show that if the anisotropic 
medium is obtained isentropically from the equilibrium one, 
then to first order in the anisotropy parameter the dissociation 
temperature remains unchanged. If, in contrast, the non-equilibrium 
system has a smaller entropy density than the equilibrium one, 
then the dissociation temperature increases with respect to 
the isotropic case, by up to $\sim$ 10\% for modest anisotropies. 

\vfill

 
\vspace*{1cm}
  
\noindent
May 2009

\vfill

\end{titlepage}

%

{\bf 1.} 
The physical picture of quarkonium dissociation within a deconfined
medium~\cite{ms} has undergone a slight refinement within the last 
couple of years. The heavy quark and anti-quark are bound together by 
almost static (off-shell) gluons, and the issue boils down to how the 
gluon self-energy looks at high temperatures. It turns out that the gluon 
self-energy has both a real and an imaginary part. The real part is 
known as Debye screening, while the imaginary part is referred to as 
Landau damping. Traditionally, it was thought that quarkonium dissociates
when Debye screening becomes so strong that the corresponding
Schr\"odinger equation supports 
no more bound state solutions. The new suggestion is that quarkonium 
effectively dissociates already at a lower temperature~\cite{static,peskin}, 
when the binding energy is non-zero but overtaken 
by the Landau-damping induced thermal width~\cite{static}. 
(Finite width phenomena have also been discussed
in a phenomenological approach~\cite{cara}.)

More quantitatively, let us denote by $M$ the heavy quark pole mass, 
by $T$ the temperature, by $r$ the quarkonium Bohr-type radius, 
and by $g$ the QCD coupling constant ($\alpha_s = g^2/4\pi$). 
Then, as is familiar from the hydrogen atom, $r\sim 1/g^2 M$. 
If quarkonium dissociates through Debye screening, the dissociation
temperature is parametrically determined by $r \mD \sim 1$, 
where $\mD \sim gT$ is the Debye mass,  leading to $T\sim g M$. 
However,  
the imaginary part of the real-time static potential~\cite{static}
overtakes the binding energy already at smaller temperature, with 
$\mD r < 1$; more precisely, in the range $g^2 M < T < g M$~\cite{peskin}. 
By formulating the problem in an effective theory framework, this estimate
was refined to $T \sim g^{4/3} M$ in ref.~\cite{es}, and further 
to $T\sim g^{4/3} \ln^{-1/3}(1/g) M$ in ref.~\cite{sewm08}. 
The last estimate was independently reproduced
with another approach in ref.~\cite{dw}. 

To summarise, the Debye-screened picture of quarkonium 
dissociation seems, in retrospective, overly conservative. 
To rephrase this more polemically, one can argue that 
Debye screening is not the 
dominant mechanism responsible for dissociating the quarkonium 
state in a thermal environment~\cite{dw}.  

All of the considerations above refer to the theoretically transparent
situation that the medium is in thermal equilibrium. 
For phenomenological applications it may be interesting to study
non-equilibrium environments as well. For instance, in ref.~\cite{lrw} 
the case was considered that the quarkonium system has a non-zero
velocity with respect to the medium. Here, in contrast, we inspect a system 
in which quarkonium is at rest but 
the ``hard'' partonic degrees of freedom have
distribution functions in momentum space which contain an anisotropy, 
or a preferred direction.
This situation could emerge as a result
of Bjorken expansion at the early stages of a central heavy ion 
collision, and it has therefore been extensively studied in the 
literature recently, for instance in the context of plasma 
instabilities~(\cite{instab} and references therein)
as well as for observables such as heavy quark energy loss \cite{rs2},
heavy quark momentum broadening~\cite{pr},
photon production~\cite{ss1}, 
dilepton production~\cite{dilepton},
jet quenching \cite{ss2,bmt}, and, most relevantly for us,   
the heavy quark potential \cite{dgs} and
solutions of the Schr\"odinger equation~\cite{dms}.  

\pagebreak


%

{\bf 2.} 
Strictly speaking, the suggestion to give the hard modes an anisotropic
momentum distribution does not lead to a self-consistent 
non-equilibrium ``ground state'': rather, the system
has ``tachyonic'' modes, meaning that one is trying to compute
around a wrong extremum. However, 
as we will see, for our observable
these problems are absent in the limit of a small anisotropy, 
and we restrict to this case in the following. 

For a general momentum distribution, the Hard Thermal Loop 
gluon self-energy obtains the form~\cite{mt}
\be
 \Pi^{\mu\nu}_R(K) = g^2 
 \int \! \frac{{\rm d}^3\vec{p}}{(2\pi)^3} \,
 v^\mu \partial^\alpha f(\vec{p}) 
 \biggl( {\delta^\nu}_\alpha - \frac{v^\nu k_\alpha}{v\cdot K + i 0^+} \biggr) 
 \;, \la{Pi_munu}
\ee
where Minkowskian conventions are assumed,
$v\equiv (1,\vec{p}/p)$, 
and the subscript $R$ indicates that the self-energy appears in the inverse
of the retarded propagator.  
Following ref.~\cite{rs1}, the hard mode momentum 
distribution function is assumed to be of the form
\be
 f(\vec{p}) \equiv f_\rmi{iso}(\sqrt{p^2 + \xi p_z^2})
 \;.  \la{ansatz}
\ee
Here $\xi \in (-1,\infty)$ is a real parameter, and
$p_z$ the momentum component in the beam direction; the values 
corresponding to the Bjorken expansion induced anisotropy are $\xi > 0$.

Carrying out a change of variables to 
$\bar{p} \equiv \sqrt{p^2 + \xi p_z^2}$, and 
denoting $p_z \equiv p \cos\theta_p$, 
the momentum integration may be written as 
\be
 \int \! \frac{{\rm d}^3\vec{p}}{(2\pi)^3} \, 
  \mathcal{F}(\vec{v}, \sqrt{p^2 + \xi p_z^2}) = 
 \int \! \frac{{\rm d}^3\bar{\vec{p}}}{(2\pi)^3} \, 
 \frac{\mathcal{F}(\vec{v},\bar{p})}{(1+\xi\cos^2\!\theta_p)^{\fr32}} 
 \;, \la{measure} 
\ee
where $\vec{v}$ is a unit vector, with $v_z = \cos\theta_p $; 
the angular variables determining the direction of $\vec{v}$ remain
unchanged in the substitution; and $\mathcal{F}$ is an arbitrary function.  

For the real-time static potential, we need 
the component $\Pi^{00}_R$ of the self-energy,  
near the static limit $|k^0| \ll k$. 
Expanding to first order in $\xi$ and $k^0$ yields
\be
 \Pi^{00}_R(K) 
 \approx \mD^2 \biggl\{ 
 -1 + \xi \biggl[ \fr16 - \fr12 \cos (2\theta_k) \biggr]
 - \frac{i \pi}{2} \frac{k_0}{k}
 \biggl[ 
 1 + \xi   \cos (2\theta_k)
 \biggr]
 \biggr\} 
 \;, \la{Pi_00}
\ee
where $\theta_k$ is the angle between $\vec{k}, \vec{z}$ and we denoted 
\be
 \mD^2 \equiv -g^2 
 \int \! \frac{{\rm d}^3\bar{\vec{p}}}{(2\pi)^3} \, 
 \frac{{\rm d}f_\rmi{iso}(\bar{p})}{{\rm d}\bar{p}}
 \;. \la{mD_def}
\ee

\vspace*{5mm}

%

{\bf 3.} 
For the real and imaginary parts of the real-time static 
potential, we need the gluon propagator near the static limit. 
More precisely, an explicit computation of the static 
potential {\em \`a la} ref.~\cite{static} can be 
rephrased by noting that, if the static limit exists
(it does at $\rmO(\xi)$ but not at $\rmO(\xi^2)$), it is the time-ordered 
gluon propagator that matters~\cite{bbr}, as would be expected 
in the naive real-time formalism:
\be
 \lim_{t\to\infty} V_{>}(t,r) = 
 g^2 C_F \int \! \frac{{\rm d}^3\vec{k}}{(2\pi)^3} \, 
 \frac{e^{i\vec{k}\cdot\vec{r}} + e^{-i\vec{k}\cdot\vec{r}} -2}{2}
 \; i \Delta_T^{00} (0,\vec{k}) 
 \;, \la{Vlarger}
\ee
where $C_F \equiv (\Nc^2 - 1)/2\Nc$ and $\Nc = 3$. 
The time-ordered propagator can in turn be written as 
\be
 i \Delta^{00}_T = \Delta^{00}_R + 2 i \nB{}(k^0) \im \Delta^{00}_R 
 \approx  \Delta^{00}_R + \frac{2 i T}{k^0} \im \Delta^{00}_R 
 \;, \la{eq7}
\ee
where $\Delta^{00}_R$ is the $00$-component of the retarded propagator, 
and $\nB{}$ is the Bose-Einstein distribution function. We have here
assumed that, unlike the hard modes, the soft gluons {\em are} already
in thermal equilibrium at a temperature $T$.  
It can be verified that at the required order in $k^0$ and $\xi$, 
the self-energies $\Pi_R^{0i}$, $\Pi_R^{i0}, \Pi_R^{ij}$ do 
not contribute to $\Delta^{00}_R$. 
Making use of \eq\nr{Pi_00} 
and expanding to first order in $\xi$, we thus obtain
that in the static limit
\ba
 i \Delta_T^{00}(0,\vec{k}) & = &  
 -\frac{1}{k^2 + \mD^2}
 + \frac{\xi \mD^2}{6 (k^2 + \mD^2)^2}
 \Bigl[3 \cos(2\theta_k) - 1 \Bigr]
 \nn & + & 
 \frac{i \pi \mD^2 T}{k (k^2 + \mD^2)^2}
 +  \frac{i \xi \pi \mD^2 T}{3 k (k^2 + \mD^2)^3}
 \Bigl[3 k^2 \cos(2\theta_k) + \mD^2 \Bigr]
 \;. \la{Delta_00} 
\ea

Inserting \eq\nr{Delta_00} into \eq\nr{Vlarger}, and noting that 
$
 \cos\theta_k = 
 \cos\theta_r \, \cos\theta_{kr} +
 \sin \theta_r\, \sin\theta_{kr} \, \cos\phi_{kr} 
$, 
where $\theta_r$ is the angle between $\vec{r}, \vec{z}$ 
and $\theta_{kr}, \phi_{kr}$ are the angular variables 
between $\vec{k}, \vec{r}$, 
we can integrate over 
$\theta_{kr}, \phi_{kr}$. 
The potential becomes
\ba
  \lim_{t\to\infty} V_{>}(t,r) & = &  
  - \frac{g^2 C_F \mD }{4\pi} \biggl\{
  \frac{e^{-\hat{r}}}{\hat{r}} + 1
  + \xi 
  \biggl[
    \frac{e^{-\hat{r}} -1}{6}
    + (1 - 3 \cos^2\!\theta_r) \rho(\hat r) 
  \biggr]
  \biggr\}
 \nn 
 & & \; - \frac{i g^2 C_F T}{4\pi} 
 \biggl\{
   \phi(\hat{r})  - \xi\biggl[ \frac{1}{3} \phi_1(\hat{r}) 
   + \frac{4}{15} (1-3 \cos^2\!\theta_r) \phi_2(\hat{r})
 \biggr]\biggr\}
 \;, \la{Vlarger2}
\ea
where $\hat r \equiv \mD r$. Furthermore, we have defined
\ba
 \rho(\hat r) & \equiv & 
    e^{-\hat{r}} \biggl( \fr16 + \fr1{2\hat{r}} + \fr1{\hat{r}^2} \biggr)
    +   \frac{e^{-\hat{r}} -1}{\hat{r}^3} 
 \;, 
 \\
 \phi(\hat r) 
 & \equiv &  2 \int_0^\infty \! \frac{{\rm d}\hat k \, \hat k} 
 {(\hat k^2 + 1)^{2}} 
 \biggl[ 1 - \frac{\sin(\hat k \hat r)}{\hat k \hat r}\biggr] 
 \;, 
 \\ 
 \phi_1(\hat r) 
 & \equiv &  2 \int_0^\infty \! \frac{ {\rm d}\hat k \, \hat k 
 (\hat k^2 - 1)}{ (\hat k^2 + 1)^{3} } 
 \biggl[ 1 - \frac{\sin(\hat k \hat r)}{ \hat k \hat r} \biggr] 
 \;,  
 \\ 
 \phi_2(\hat r) 
 & \equiv &  5 \int_0^\infty \! \frac{{\rm d}\hat k \, \hat k^3 }{
 (\hat k^2 + 1)^{3} } 
 \biggl[ \frac{ 3 \sin(\hat k \hat r)}{ (\hat k \hat r)^3 } - 
   \frac{3 \cos(\hat k \hat r)}{(\hat k \hat r)^2} -
   \frac{\sin(\hat k \hat r)}{\hat k \hat r } \biggr] 
 \;.
\ea
The real part of the potential agrees with the result of ref.~\cite{dgs}.
The functions appearing in the imaginary part are finite for all $\hat{r}$; 
$\phi_1$ and $\phi_2$ vanish both at small and large distances. 

\vspace*{5mm}

%

{\bf 4.} 
In order to obtain a theoretically consistent result in the weak-coupling
regime, it is important to systematically account for all the effects to 
a given order, emerging from 
the various momentum and frequency scales of the problem. 
This can be achieved by employing effective field theory 
methods~\cite{es,nb3} (for a review, see ref.~\cite{sewm08}). The upshot is 
that $\hat{r}$ is formally a small parameter, $\hat{r} < 1$~\cite{peskin}, 
and can be expanded in.

Expanding in $\hat{r}$, the term proportional to $\xi$ of the real
part of the potential is seen to be of $\rmO(g^2 \mD^2 r)$, i.e.\ 
suppressed by two powers of $\hat r$ with respect to the $\xi$-independent
term. Therefore it can be omitted in the weak-coupling limit. 

In contrast, in the imaginary part (the second line of \eq\nr{Vlarger2})
the corrections from $\xi$ are of the same order as the leading term: 
\ba
 \phi(\hat{r}) & \approx & 
 \frac{\hat{r}^2}{3} \biggl( 
   \ln\frac{1}{\hat{r}} - \gamma_\rmii{E} + \fr43 
 \biggr)
 \;, \\ 
 \phi_1(\hat{r}) & \approx & 
 \frac{\hat{r}^2}{3} \biggl( 
   \ln\frac{1}{\hat{r}} - \gamma_\rmii{E} + \fr56 
 \biggr)
 \;, \\ 
 \phi_2(\hat{r}) & \approx & 
 \frac{\hat{r}^2}{3} \biggl( 
   \ln\frac{1}{\hat{r}} - \gamma_\rmii{E} + \frac{47}{60} 
 \biggr)
 \;. 
\ea
In particular, at leading logarithmic order, all three functions
behave identically.

\vspace*{5mm}

%

{\bf 5.} 
Before we inspect more precisely the consequences of \eq\nr{Vlarger2}, 
we need to discuss the value of the parameter $\mD$ appearing in it, 
defined by \eq\nr{mD_def}.

Often the function $f_\rmi{iso}$ in  \eq\nr{mD_def}
is taken to be a sum of Bose-Einstein
and Fermi-Dirac distribution functions, with a ``temperature'' $T$ appearing
as a parameter (see, e.g., eq.~(1) of ref.~\cite{cr}). 
It is clear, however, that since $f_\rmi{iso}$
was part of a specification of a non-equilibrium state (cf.\ \eq\nr{ansatz}),
a precise physical meaning can only be given to $T$ through further arguments 
(cf.\ ref.~\cite{akr} and references therein).
We have already specified our ``late-time'' setting above 
(strongly interacting 
soft modes thermalized at a {\em physical} temperature $T$, 
weakly interacting hard modes still anisotropic), 
so we need to rethink the meaning of the 
parameter appearing in $f_\rmi{iso}$ for this case. 
We denote this parameter by $T'$ in the following.  

Consider now the entropy density
of the system. Making use of the substitution 
in \eq\nr{measure}, it is given by an integral of the type 
\be
  s_\rmi{non-eq}
 = \int \! \frac{{\rm d}^3\vec{p}}{(2\pi)^3} \, 
  \mathcal{F}(\bar{p}) \approx 
 \int \! \frac{{\rm d}^3\bar{\vec{p}}}{(2\pi)^3} \, 
 \mathcal{F}(\bar{p}) \Bigl(1-\fr32 \xi\cos^2\!\theta_p\Bigr) 
 \;, 
\ee
where we expanded to leading order in $\xi$. Carrying out the angular
integral, and assuming a massless system without chemical potentials, 
this yields 
\be 
 s_\rmi{non-eq} = \Bigl( 1 - \frac{\xi}{2}\Bigr)\, c \, T'^3
 \;, 
\ee
with a certain constant $c$.
Similarly, the energy density becomes 
$
 e_\rmi{non-eq} = \fr34 (1-2\xi/3) c T'^4
$.

We could now envisage two cases. If we {\em define} the non-equilibrium
state by $T' \equiv T$, then we observe that, for $\xi > 0$, 
it has less entropy density than 
the corresponding equilibrium state, with $s_\rmi{eq} = c\, T^3$. On the other
hand, if we impose the physical condition that the non-equilibrium state
be related ``isentropically'' to the equilibrium one, then we are 
left to conclude that the parameter $T'$ should be chosen as 
\be
 T' = \Bigl( 1 + \frac{\xi}{6} \Bigr) T
 \;. \la{Tp}
\ee
The same outcome results if we define 
the temperature through $T^{-1} \equiv 
\partial s_\rmi{non-eq}/\partial e_\rmi{non-eq}$.

\vspace*{5mm}

%

{\bf 6.} 
In order to estimate the temperature at which quarkonium dissociates, we 
finally carry out a parametric computation according to the discussion above. 
As we will see the effects of the anisotropy parameter $\xi$ can be 
fully accounted for without being concerned about various numerical 
factors. Thus, we estimate the magnitude of the binding energy by the 
real part of the static potential, expanded to leading order in $\hat r$
and evaluated at the distance scale of the Bohr radius: 
\be
 \re [V_>] \sim - \frac{g^2 C_F}{4\pi r} 
 \left. \Bigl[ 1 + \rmO(\hat{r}) \Bigr] \right|_{r\sim 1/g^2M}
 \;. \la{reVg}
\ee
In the imaginary part, in contrast, corrections involving $\xi$ are of
order unity. Considering the $s$-wave ground state, the term proportional 
to $1-3 \cos^2\!\theta_r$ in \eq\nr{Vlarger2} averages to zero
at first order in perturbation theory
(corrections will be of order $\xi^2$), 
so to leading-logarithmic order 
\be
 \im [V_>] \sim -\frac{g^2 C_F T}{4\pi} \frac{\mD^2 r^2}{3} 
 \ln\frac{1}{\mD r} 
 \Bigl[ 1 + \rmO\Bigl(\frac{1}{\ln\hat{r}}\Bigr) \Bigr]
 \times \Bigl( 1 - \frac{\xi}{3} \Bigr) 
 \;. \la{imVg}
\ee
If we now assume the temperature $T'$ parameterizing the non-equilibrium
system to be isentropically obtained from the equilibrium case, leading 
to \eq\nr{Tp}, then the Debye mass parameter defined in \eq\nr{mD_def}
evaluates to 
\be
 \mD^2 \sim g^2 T'^2 \approx g^2 T^2 \Bigl( 1 + \frac{\xi}{3} \Bigr)
 \;. 
\ee
Thus $\xi$-dependence cancels on the right-hand side of \eq\nr{imVg}
to $\rmO(\xi)$; at leading-logarithmic order 
the dissociation temperature remains at the value 
\be
 T_\rmi{melt} \sim 
 g^{\fr43} (\ln{\textstyle\frac{1}{g}} )^{-\fr13} M \;,
 \la{Tmelt1} 
\ee
obtained in refs.~\cite{sewm08,dw}.

If, in contrast, we assume $T' = T$, then $\mD^2 \sim g^2 T^2$. 
Temperature appears in \eq\nr{imVg} cubically, 
meaning that equality with the 
real part of \eq\nr{reVg} is obtained for 
\be
 T_\rmi{melt} \sim 
 g^{\fr43} (\ln{\textstyle\frac{1}{g}} )^{-\fr13} M
 \times \biggl( 1 + \frac{\xi}{9} \biggr) \;. \la{Tmelt2}
\ee
For $\xi \sim 1$, up to which range our small-$\xi$ approach might
be assumed qualitatively reasonable, we thus obtain a 10\% increase
in the dissociation temperature. 

\vspace*{5mm}

%

{\bf 7.} 
To conclude, while the precise definition 
of the heavy quarkonium dissociation temperature is 
ambiguous, requiring a convention on the shape of the corresponding
smoothly evolving spectral function (for results within the weak-coupling 
expansion, see ref.~\cite{nlo}), it appears that for the class
of non-equilibrium states introduced in ref.~\cite{rs1}, the 
{\em relative change} caused
by a momentum space anisotropy can be estimated analytically
at leading-logarithmic order. 
On the other hand, the definition of the non-equilibrium
state itself contains a hidden ambiguity, 
in that a temperature-like parameter is introduced which requires 
further justification. We have related this parameter to 
the entropy density of the system, and thus arrived at
two physically distinct results, \eqs\nr{Tmelt1} and \nr{Tmelt2}. 
In general, it is expected that non-equilibrium states have 
less entropy than the equilibrium one, under which conditions 
\eq\nr{Tmelt2} could be a better estimate than \eq\nr{Tmelt1}, indicating
an increase of the dissociation temperature; nevertheless, reasonable
arguments could also be given in favour of \eq\nr{Tp}, 
leading to \eq\nr{Tmelt1}. 
In any case, our results
suggest that quarkonium dissociation is primarily 
a probe of the entropy density of the system, 
i.e.\ of the number of hard modes, 
rather than of Debye screening, i.e.\ of soft collective phenomena.   
At the same time, given the ambiguities 
appearing, it seemed to us that the analytic leading-logarithmic
order-of-magnitude estimate is about as far as one needs to go for small 
anisotropies; for larger ones, numerical simulations would be needed.

%
\section*{Note added}

Recently a paper appeared~\cite{ds}
where the same problem is considered as here. 
The results deviate slightly from ours because the physics 
setting is different: unlike in our \eq\nr{eq7},
even the soft gluons are assumed to have 
a non-thermal distribution function, 
which leads to an additional 
term in the gluon propagator (the 2nd line of eq.~(14) in ref.~\cite{ds}). 
We note, however, that if we re-express within our setting 
the soft gluon $T$ in terms of $T'$ from \eq\nr{Tp}, 
and identify $T'$ with the temperature parameter of ref.~\cite{ds}, 
then curiously our \eq\nr{imVg} does reproduce eq.~(58) of ref.~\cite{ds}.

%
\section*{Acknowledgements}

M.L. thanks A.~Dumitru, M.~Martinez and A.~Rebhan for useful correspondence. 
We are grateful to the BMBF for financial support under project
{\em Hot Nuclear Matter from Heavy Ion Collisions 
     and its Understanding from QCD}.
M.V.\ was partly supported by the Academy of Finland, 
contract no.\ 128792.


\appendix
\renewcommand{\thesection}{Appendix~\Alph{section}}
\renewcommand{\thesubsection}{\Alph{section}.\arabic{subsection}}
\renewcommand{\theequation}{\Alph{section}.\arabic{equation}}


\end{document}